\documentclass{article}

\PassOptionsToPackage{numbers, compress}{natbib}

     \usepackage[preprint]{neurips_2019}

\usepackage[utf8]{inputenc} 
\usepackage[T1]{fontenc}    
\usepackage{hyperref}       
\usepackage{url}            
\usepackage{booktabs}       
\usepackage{amsfonts}       
\usepackage{nicefrac}       
\usepackage{microtype}      
\usepackage{wrapfig}
\usepackage{microtype}
\usepackage{graphicx}
\usepackage{subfigure}
\usepackage{booktabs} 
\usepackage{amsmath}
\usepackage{enumitem}
\usepackage{framed}

\usepackage{xstring}
\newif\ifupper\uppertrue
\newcommand{\mixedcase}[1]{%
\StrSplit{#1}{1}{\currentchr}{\tailchr}%
\ifupper\MakeUppercase{\currentchr}\upperfalse\else\MakeLowercase{\currentchr}\uppertrue\fi%
\IfStrEq{\tailchr}{}{\uppertrue}{\mixedcase{\tailchr}}%
}

\usepackage{hyperref}
\usepackage{listings}
\usepackage[disable]{todonotes}
\usepackage{dsfont}

\newcommand{\eval}[1]{[\![{#1}]\!]}

\usepackage{amsmath,amssymb}
\DeclareMathOperator{\E}{\mathbb{E}}

\newcommand{\1}[1]{\mathds{1}\left[#1\right]}
\usepackage{ wasysym }
\usepackage{booktabs}
\title{Write, Execute, Assess: \\ Program Synthesis with a REPL}

\author{Kevin Ellis\thanks{These authors contributed equally to this work.}\\MIT
  \And
  Maxwell Nye$^*$\\
  MIT
  \And
  Yewen Pu$^*$\\
  MIT
  \And
  Felix Sosa$^*$\\
  Harvard University
  \AND
  Joshua B. Tenenbaum\\
  MIT
  \And
  Armando Solar-Lezama\\
  MIT}

\begin{document}
\maketitle
\begin{abstract}
  We present a neural program synthesis approach integrating components which write, execute, and assess code to navigate the search space of possible programs. We equip the search process with an interpreter or a read-eval-print-loop (REPL), which immediately executes partially written programs, exposing their semantics. The REPL addresses a basic challenge of program synthesis: tiny changes in syntax can lead to huge changes in semantics. We train a pair of models, a policy that proposes the new piece of code to write, and a value function that assesses the prospects of the code written so-far. At test time we can combine these models with a Sequential Monte Carlo algorithm. We apply our approach to two domains: synthesizing text editing programs and inferring 2D and 3D graphics programs.


\end{abstract}

\section{Introduction}

When was the last time you typed out a large body of code all at once, and had it work on the first try? If you are like most programmers, this hasn't happened much since ``Hello, World.'' Writing a large body of code is a process of \emph{trial and error} that alternates between trying out a piece of code, executing it to see if you're still on the right track, and trying out something different if the current execution looks buggy. Crucial to this human work-flow is the ability to \emph{execute} the partially-written code, and the ability to \emph{assess} the resulting execution to see if one should continue with the current approach. Thus, if we wish to build machines that automatically write large, complex programs, designing systems with the ability to effectively transition between states of writing, executing, and assessing the code may prove crucial. 

In this work, we present a model that integrates components which write, execute, and assess code to perform a stochastic search over the \emph{semantic} space of possible programs. 
We do this by equipping our model with one of the oldest and most basic tools available to a programmer: an interpreter, or read-eval-print-loop (REPL), which immediately executes partially written programs, exposing their semantics. The REPL addresses a fundamental challenge of program synthesis: tiny changes in syntax can lead to huge changes in semantics. The mapping between syntax and semantics is a difficult relation for a neural network to learn, but comes for free given a REPL. By conditioning the search solely on the execution states rather than the program syntax, the search is performed entirely in the \emph{semantic} space. By allowing the search to branch when it is uncertain, discard branches when they appear less promising, and attempt more promising alternative candidates, the search process emulates the natural iterative coding process used by a human programmer.

\todo{need better transition}In the spirit of systems such as AlphaGo~\citep{alphaGo}, we train a pair of models -- a \emph{policy} that proposes new pieces of code to write, and a \emph{value function} that evaluates the long-term prospects of the code written so far, and deploy both at test time in a symbolic tree search. Specifically, we combine the policy, value, and REPL with a Sequential Monte Carlo (SMC) search strategy at inference time. We sample next actions using our learned policy, execute the partial programs with the REPL, and re-weight the candidates by the value of the resulting partial program state. This algorithm allows us to naturally incorporate writing, executing, and assessing partial programs into our search strategy, while managing a large space of alternative program candidates.

Integrating learning and search to tackle the problem of program
synthesis is an old idea experiencing a recent
resurgence~\cite{ganin2018synthesizing,schmidhuber2004optimal,ellis2018learning,balog2016deepcoder,nye2019learning,TikZ,devlin2017robustfill,kalyan2018neural,tian2019learning,pu2018selecting}.
Our work builds on recent ideas termed `execution-guided neural
program synthesis,' independently proposed by
\citep{chen2018execution} and~\citep{zohar2018automatic}, where a
neural network writes a program conditioned on
intermediate execution states.  We extend these ideas along two
dimensions. First, we cast these different kinds of execution guidance
in terms of interaction with a REPL, and use reinforcement learning
techniques to train an agent to both interact with a REPL, \emph{and}
to assess when it is on the right track. Prior execution-guided neural synthesizers do not learn to \emph{assess} the execution state,
which is a prerequisite for sophisticated search algorithms,
like those we explore in this work. \todo{moved these sentences around per Armando request. maybe cut second half?} Second, we investigate
several ways of interleaving the policy and value networks
during search, finding that an SMC sampler provides an effective
foundation for an agent that writes, executes and assesses its code.

\begin{figure*}
  \centering
    \includegraphics[width=13.75cm,height=8cm,keepaspectratio]{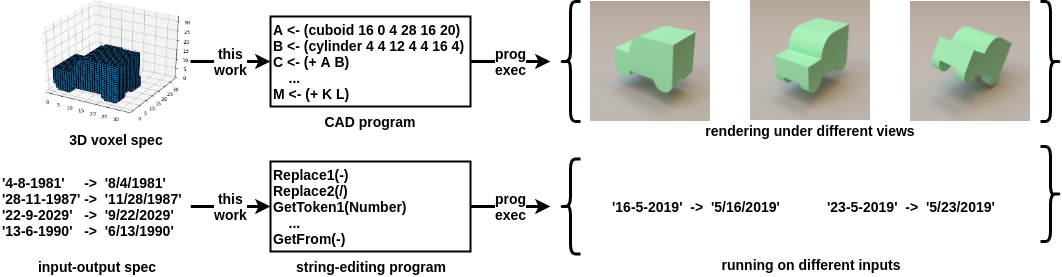}
  \caption{Examples of programs synthesized by our system. Top, graphics program from voxel specification. Bottom, string editing program from input-output specification.}
    \label{fig:we-are-cool}
\end{figure*}

We validate our framework on two different domains (see Figure~\ref{fig:we-are-cool}): inferring 2D and 3D graphics programs (in the style of computer aided design, or CAD) and synthesizing text-editing programs (in the style of FlashFill~\cite{gulwani2011automating}). In both cases we show that a code-writing agent equipped with a REPL and a value function to guide the search achieves faster, more reliable program synthesis.

Explicitly, our contributions are:
\begin{itemize}[leftmargin=*]
    \item We cast the problem of program synthesis as a search problem and solve it using Sequential Monte Carlo (SMC) by employing a pair of learned functions: a value and a policy. \todo{change to REPL and search in semantic space only}
    \item We equip the model with a REPL, bridging the gap between the syntax and semantics of a program.
    \item We empirically validate our technique in two different program synthesis domains, text-editing programs and 2D/3D graphics programs, and 
      out-perform alternative approaches.
\end{itemize}

\section{An Illustrative Example}

\begin{figure*}
  \centering
    \includegraphics[width=\textwidth,keepaspectratio]{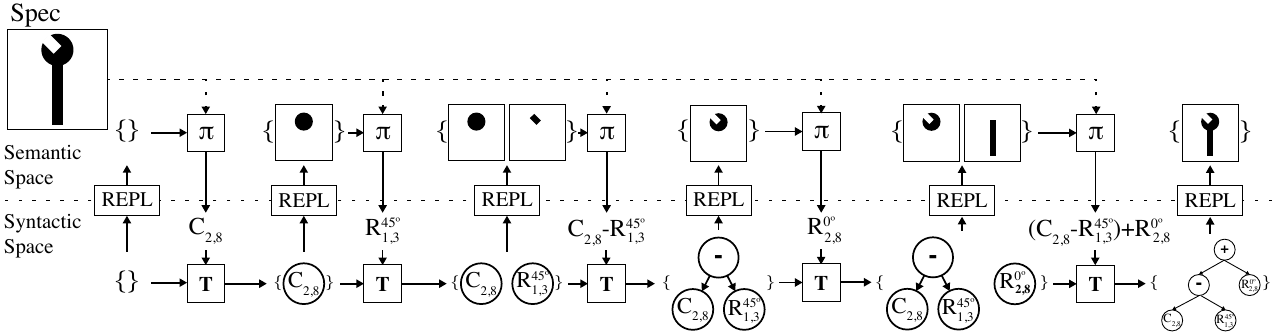}
  \caption{A particular trajectory of the policy building a 2D wrench. At each step, the REPL renders the set of partial programs $pp$ into the semantic (image) space. These images are fed into the policy $\pi$ which proposes how to extend the program via an action $a$, which is incorporated into $pp$ via the transition $T$.}
    \label{fig:pac_wave}
\end{figure*}

To make our framework concrete, consider the following program synthesis task of synthesizing a constructive solid geometry (CSG) representation of a simple 2D scene (see Figure~\ref{fig:pac_wave}). CSG is a shape-modeling language that allows the user to create complex renders by combining simple primitive shapes via boolean operators. The CSG program in our example consists of two boolean combinations: union $+$ and subtraction $-$ and two primitives: circles $C_{x,y}^{r}$ and rectangles $R_{x,y}^{w,h,\theta}$, specified by position $x,y$, radius $r$, width and height $w,h$, and rotation $\theta$. This CSG language can be formalized as the context-free grammar below: 
\begin{lstlisting}[mathescape=true]
P -> P $+$ P | P $-$ P | $C_{x,y}^r$ | $R_{x,y}^{w,h,\theta}$
\end{lstlisting}

The synthesis task is to find a CSG program that renders to $spec$.
Our policy constructs this program one piece at a time, conditioned on the set of expressions currently in scope. Starting with an empty \emph{set} of programs in scope, $pp = \{\}$, the policy proposes an action $a$ that extends it. This proposal process is iterated to incrementally extend $pp$ to contain longer and more complex programs. In this CSG example, the action $a$ is either adding a primitive shape, such as a circle $C_{2,8}^3$, or applying a boolean combinator, such as $p_1-p_2$, where the action also specifies its two arguments $p_1$ and $p_2$.

To help the policy make good proposals, we augment it with a REPL, which takes a set of programs $pp$ in scope and executes each of them. In our CSG example, the REPL renders the set of programs $pp$ to a set of images. The policy then takes in the REPL state (a set of images), along with the specification $spec$ to predict the next action $a$. This way, the input to the policy lies entirely in the semantic space, akin to how one would use a REPL to iteratively construct a working code snippet. Figure~\ref{fig:pac_wave} demonstrates a potential roll-out through a CSG problem using only the policy.
\label{sec:example}

\begin{figure*}
  \centering
    \includegraphics[width=0.8\textwidth,keepaspectratio]{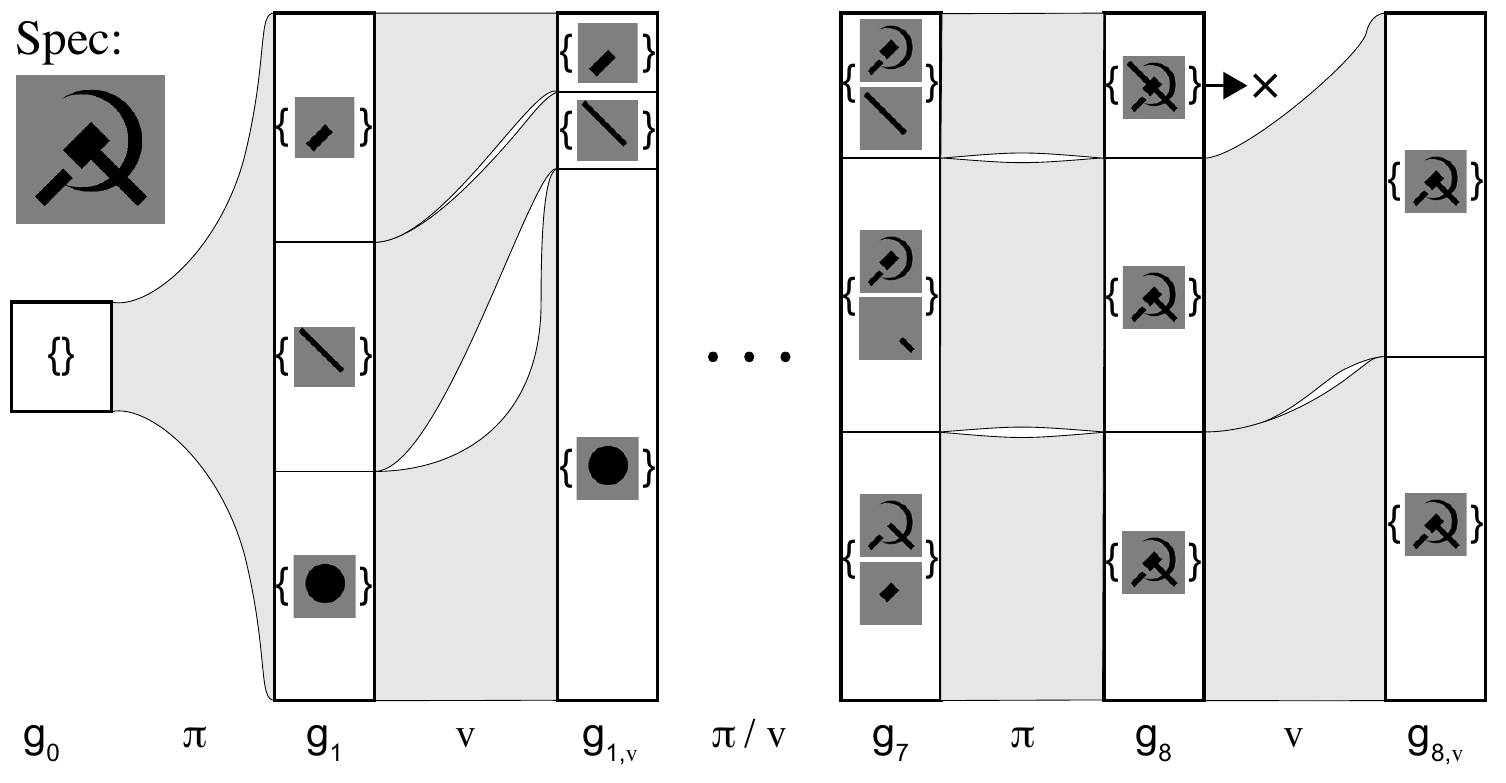}
  \caption{SMC sampler maintains a population of particles (i.e. programs), which it evolves forward in time by (1) sampling from policy $\pi$ to get a new generation of particles, then (2) reweighting by the value function $v$ and resampling to prune unpromising candidates and up-weight promising ones.}
    \label{fig:csg_search}
\end{figure*}

However, code is brittle, and if the policy predicts an incorrect action, the entire program synthesis fails. To combat this brittleness, we use Sequential Monte Carlo (SMC) to search over the space of candidate programs. Crucial to our SMC algorithm is a learned value function $v$ which, given a REPL state, assesses the likelihood of success on this particular search branch. By employing $v$, the search can be judicious about which search branch to prioritize in exploring and withdraw from branches deemed unpromising. Figure~\ref{fig:csg_search} demonstrates a fraction of the search space leading up to the successful program and how the value function $v$ helps to prune out unpromising search candidates.

\section{Our Approach}

Our program synthesis approach integrates components that write, execute, and assess code to perform a stochastic search over the semantic space of programs. Crucial to this approach are three main components: First, the definition of the search space; Second, the training of the code-writing policy and the code-assessing value function; Third, the Sequential Monte Carlo algorithm that leverages the policy and the value to conduct the search by maintaining a population of candidate search branches.



\subsection{The Semantic Search Space of Programs}
The space of possible programs is typically defined by a context free grammar (CFG), which specifies the set of syntactically valid programs. However, when one is writing the code, the programs are often constructed in a piece-wise fashion. Thus, it is natural to express the search space of programs as a markov decision process (MDP) over the set of partially constructed programs. 

\noindent\textbf{State}\quad
The state is a tuple $s = (pp, spec)$ where $pp$ is a \emph{set} of partially-constructed program trees (intuitively, `variables in scope'), and $spec$ is the goal specification. Thus, our MDP is goal conditioned. The start state is $(\{\}, spec)$. 

\noindent\textbf{Action}\quad
The action $a$ is a production rule from the CFG (a line of code typed into the REPL).

\noindent\textbf{Transitions}\quad
The transition, $T$, takes the set of partial programs $pp$ and applies the action $a$ to either:\\
1.\quad instantiate a new sub-tree if $a$ is a terminal production: $T(pp,a) = pp \cup \{a\}$\\
2.\quad combine multiple sub-trees if $a$ is a non-terminal: $T(pp,a) = (pp\cup \{a(t_1 \dots t_k)\}) - \{t_1 \dots t_k\}$

Note that in the case of a non-terminal, the children $t_1 \dots t_k$ are removed, or `garbage-collected'~\cite{zohar2018automatic}.

\noindent\textbf{Reward}\quad
The reward is $1$ if there is a program $p \in pp$ that satisfies the spec, and $0$ otherwise.

Note that the state of our MDP is defined jointly in the syntactic space, $pp$, and the semantic space, $spec$. To bridge this gap, we use a REPL, which evaluates the set of partial programs $pp$ into a semantic or ``executed'' representation. Let $pp$ be a set of $n$ programs, $pp = \{p_1 \dots p_n\}$ and let $\eval{p}$ denote the execution of a program $p$, then we can write the REPL state as $\eval{pp} = \{\eval{p_1} \dots \eval{p_n}\}$.

\subsection{Training the Code-Writing Policy $\pi$ and the Code-Assessing Value $v$}

Given the pair of evaluated program states and spec ($\eval{pp}$, $spec$), the policy $\pi$ outputs a distribution over actions, written $\pi(a ~|~ \eval{pp}, spec)$, and the value function $v$ predicts the expected reward starting from state ($\eval{pp}$, $spec$). In our MDP, expected total reward conveniently coincides with the probability of a rollout satisfying $spec$ starting with the partial program $pp$:
\begin{align}
  v\left( \eval{pp}, spec\right) &= \E_\pi\left[R~|~\eval{pp}, spec \right] = \E_\pi\left[\1{spec\text{ is satisfied}}~|~\eval{pp}, spec \right]\nonumber\\ &= \text{P}(\text{rollout w/ $\pi$ satisfies $spec$}~|~\eval{pp}, spec)
\end{align}
Thus the value function simply performs binary classification, predicting the probability that a state will lead to a successful program.

\textbf{Pretraining $\pi$.}\quad
Because we assume the existence of a CFG and a REPL, we can generate an infinite stream of training data by sampling random programs from the CFG, executing them to obtain a spec, and then recovering the ground-truth action sequence. Specifically, we draw samples from a distribution over synthetic training data, $\mathcal{D}$, consisting of triples of the spec, the sequence of actions, and the set of partially constructed programs at each step: $(spec, \left\{a_t \right\}_{t\leq T}, \left\{pp_t \right\}_{t\leq  T}) \sim \mathcal{D}$. During pretraining, we maximize the log likelihood of these action sequences under the policy:
\begin{equation}
  \mathcal{L}^{\text{pretrain}}(\pi) = \E_{\mathcal{D}}\left[\sum_{t\leq T}\log \pi(a_t|\eval{pp_t},spec) \right]
\end{equation}
\textbf{Training $\pi$ and $v$.}\quad
We fine-tune the policy and train the value function by sampling the policy's roll-outs against  $spec\sim\mathcal{D}$ in the style of REINFORCE. Specifically, given $spec \sim \mathcal{D}$, the policy's rollout consists of a sequence of actions $\left\{a_t \right\}_{t\leq T}$, a sequence of partial programs $\left\{pp_t \right\}_{t\leq T}$, and a reward $R = \mathds{1}\left[\eval{pp_T}\text{ satisfies spec} \right]$. Given the specific rollout, we train $v$ and $\pi$ to maximize:
\begin{align}
  \mathcal{L}^{\text{RL}}(v,\pi) = ~ &R  \sum_{t\leq T}\log v(\eval{pp_t},spec) + (1 - R) \sum_{t\leq T}\log (1 - v(\eval{pp_t},spec)) \nonumber\\+ &R\sum_{t\leq T}\log \pi(a_t|\eval{pp_t},spec)
\end{align}


\subsection{An SMC Inference Algorithm That Interleaves Writing, Executing, and Assessing Code}
At test time we interleave \emph{code writing}, i.e. drawing actions from the policy, and \emph{code assessing}, i.e. querying the value function (and thus also interleave execution, which always occurs before running these networks). \todo{almost there but a little awkward}
Sequential Monte Carlo methods~\cite{doucet2001introduction}, of which particle filters are the most famous example, are a flexible framework for designing samplers that infer a sequence of $T$ latent variables $\left\{x_t \right\}_{t\leq T}$ conditioned on a paired sequence of observed variables $\left\{y_t \right\}_{t\leq T}$. Following~\cite{ellis2018learning} we construct an SMC sampler for our setting by identifying the policy rollout over time as the sequence of latent variables (i.e. $x_t = pp_t$),
and identify the spec as the observed variable at every time step (i.e. $y_t = spec$),
which are connected to the policy and value networks by defining
\begin{align}
  \text{P}(x_{t + 1}|x_t) = \sum_{\substack{a\\T(x_t,a) = x_{t + 1}}}\pi(a|x_t)\qquad&&\text{P}(y_t|x_t) \propto v(\eval{x_t},y_t)
\end{align}
and, like a particle filter, we approximately sample from $\text{P}(\left\{pp_t \right\}_{t\leq T}|spec)$ by maintaining a population of $K$ particles -- each particle a state in the MDP -- and evolve this population of particles forward in time by sampling from $\pi$, importance reweighting by $v$, and then resampling.

SMC techniques are not the only reasonable approach:
one could perform a beam search, seeking to maximize $\log \pi(\left\{a_t \right\}_{t < T}|spec) + \log v(\eval{pp_T},spec)$;
or, A* search by interpreting $-\log \pi(\left\{a_t \right\}_{t < T}|spec)$ as cost-so-far and $-\log v(\eval{pp_T},spec)$ as heuristic cost-to-go;
or, as popularized by AlphaGo, one could use MCTS jointly with the learned value and policy networks.
SMC confers two main benefits:
(1) it is a stochastic search procedure,
immediately yielding a simple any-time algorithm where we repeatedly run the sampler and keep the best program found so far;
and (2) the sampling/resampling steps are easily batched on a GPU,
giving high throughput unattainable with serial algorithms like A*.

\section{Experiments}

\begin{figure*}[b]
  \centering
    \includegraphics[width=14cm,height=8cm,keepaspectratio]{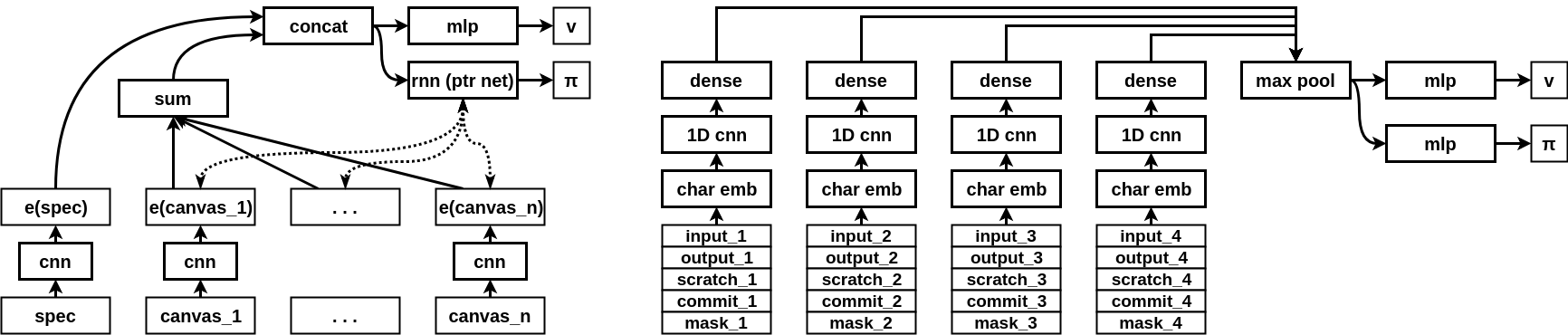}
   \caption{Left: CAD architecture. The policy is a CNN followed by a pointer network~\cite{vinyals2015pointer} (attending over the set of partial programs $pp$)  which decodes into the next line of code.
The value function is an additional `head' to the pooled CNN activations.
Right: String Editing architecture. We encode each example using an embedding layer, apply a 1-d convolution and a dense block, and pool the example hidden states to predict the policy and value.}
    \label{fig:architecture}
\end{figure*}

To assess the relative importance of the policy, value function, and REPL,
we study a spectrum of models and test-time inference strategies in both of our domains.
For each model and inference strategy, we are interested in how efficiently it can search the space of programs,
i.e. the best program found as a function of time spent searching.
We trained a pair of models: our REPL model, which conditions on intermediate execution states (architectures in Figure~\ref{fig:architecture}), and a `no REPL' baseline, which decodes a program in one shot using only the spec and syntax. This baseline is inspired by the prior work CSGNet~\cite{sharma2018csgnet} and RobustFill~\cite{devlin2017robustfill} for CAD and string editing, respectively.
We compare our SMC sampler with a simple beam search using both policy and value, as well as a pair of inference strategies employing only the policy: simple policy rollouts, and a beam search decoding to find the highest likelihood program under $\pi$. The policy-only beam search method is inspired by~\cite{chen2018execution} and~\cite{zohar2018automatic}, which performed execution-guided synthesis in Karel and list processing domains, respectively.

\subsection{Inverse CAD}

Modern mechanical parts are created using Computer Aided Design (CAD),
a family of programmatic shape-modeling techniques.
Here we consider two varieties of \emph{inverse} CAD: inferring programs generating 3D shapes, and programs generating 2D graphics,
which can also be seen as a kind of high-level physical object understanding in terms of parts
and relations between parts.

We use CSG as our CAD modeling language,
where
the specification $spec$ is an image, i.e. a pixel/voxel array for 2D/3D, 
and the the goal is to write a program that renders to the target image.
These programs build shapes by algebraically combining primitive drawing commands via addition and subtraction, including circles and rotated rectangles (for 2D, w/ coordinates quantized to $16\times 16$) as well as spheres, cubes, and cylinders (for 3D, w/ coordinates quantized to $8\times 8\times 8$),
although in both cases the quantization could in principle be made arbitrarily fine.
Our REPL renders each partial program $p \in pp$ to a distinct canvas, which the policy and value networks take as input.

\begin{figure*}[b!]
  \setlength{\tabcolsep}{0pt}
  \renewcommand{\arraystretch}{0}
  \centering  \begin{tabular}{ccc}
    \begin{tabular}{c}
      \includegraphics[width = 0.4\textwidth]{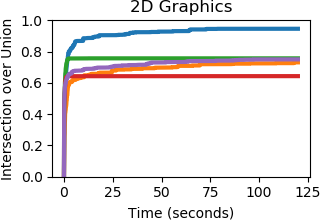}
    \end{tabular}&
    \begin{tabular}{c}
$\quad$      \includegraphics[width = 0.4\textwidth]{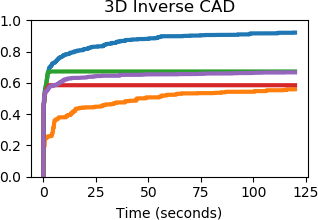}
    \end{tabular}&
    \begin{tabular}{c}
      \includegraphics[width = 0.15\textwidth]{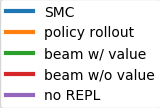}
    \end{tabular}
  \end{tabular}
  \setlength{\tabcolsep}{6pt}
  \renewcommand{\arraystretch}{1}
  \caption{Quantitative results for CAD on out-of-sample testing problems. Both models trained on scenes with up to $13$ objects. Left: 2D models tested on scenes with up to $30$ objects. Right: 3D models tested on scenes with up to $20$ objects. SMC achieves the highest test accuracy.}  \label{graphics_quantitative}
\end{figure*}

\begin{figure}
  \setlength{\tabcolsep}{0pt}
  \begin{tabular}{c}
  
      \includegraphics[width=\textwidth,keepaspectratio]{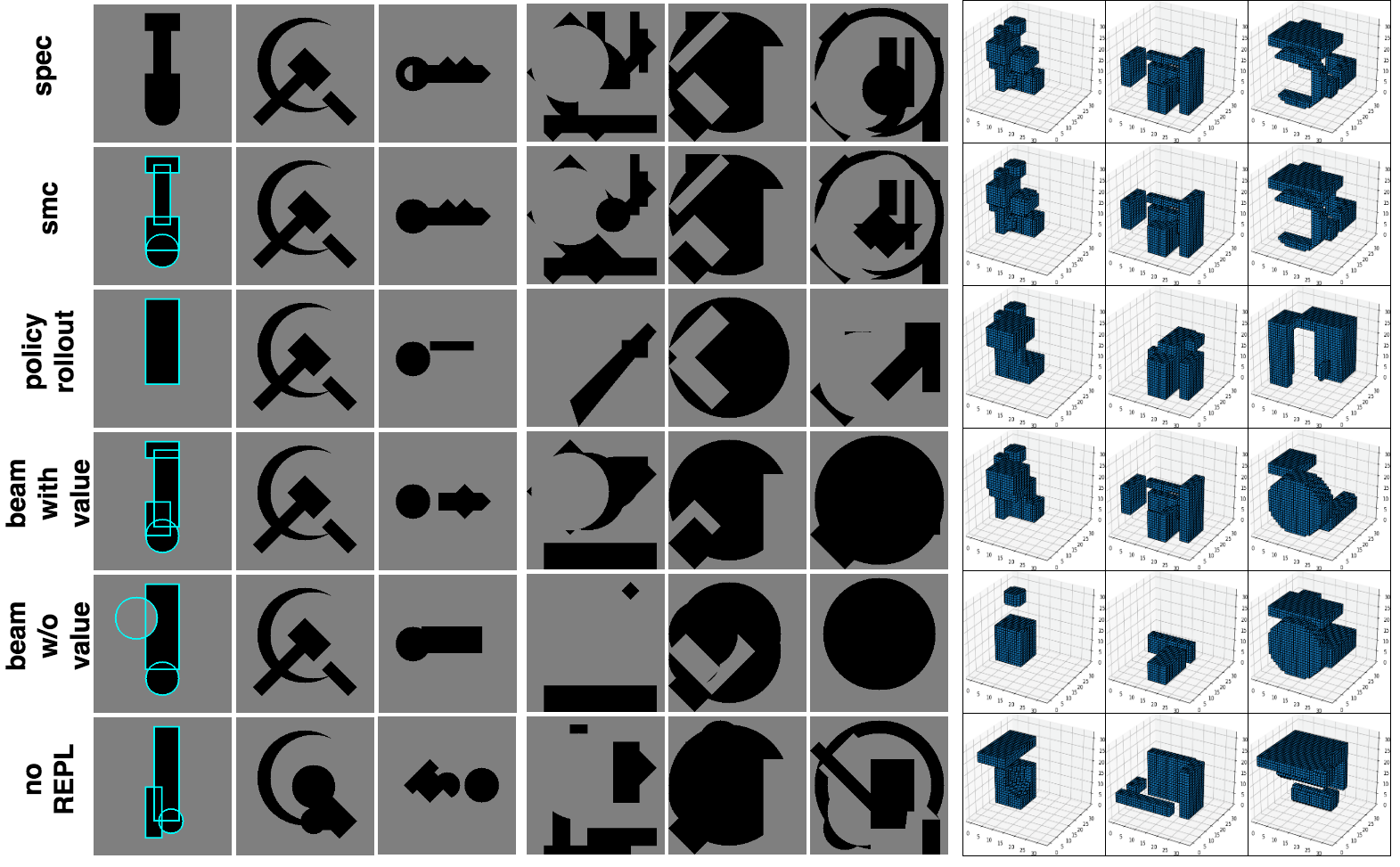} \\\\
  \renewcommand{\arraystretch}{0}  
  
  \begin{tabular}{cccccccc}
    \rotatebox[origin=l]{90}{Spec}
    &    \rotatebox[origin=l]{90}{(voxels)}$\;$&
      \includegraphics[width = 2cm]{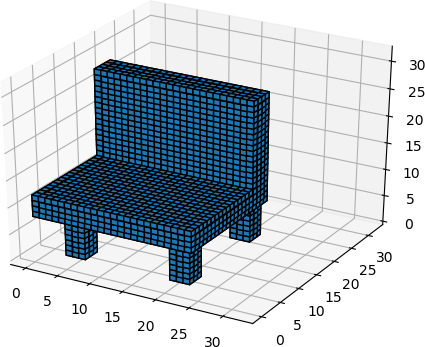}&
      \includegraphics[width = 2cm]{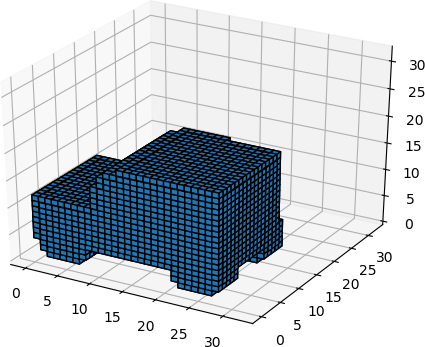}&
      \includegraphics[width = 2cm]{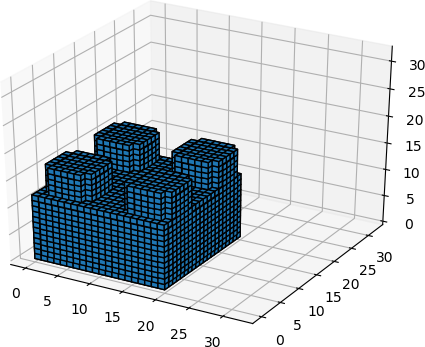}&
      \includegraphics[width = 2cm]{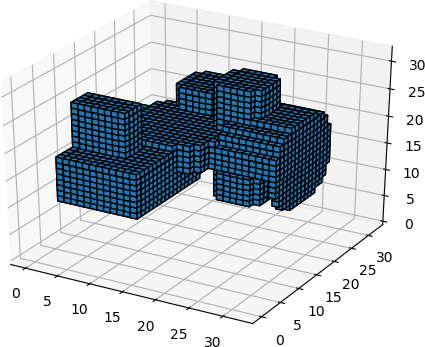}&
      \includegraphics[width = 2cm]{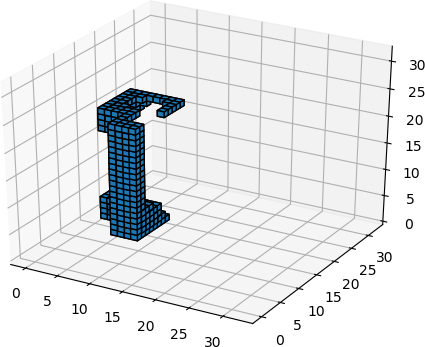}&
      \includegraphics[width = 2cm]{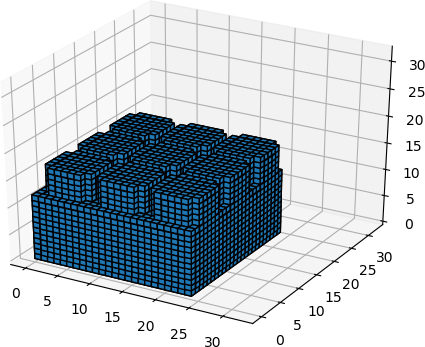}
      \\
      \rotatebox[origin=l]{90}{Rendered}&
      \rotatebox[origin=l]{90}{program}$\;$&
      \includegraphics[width = 2cm]{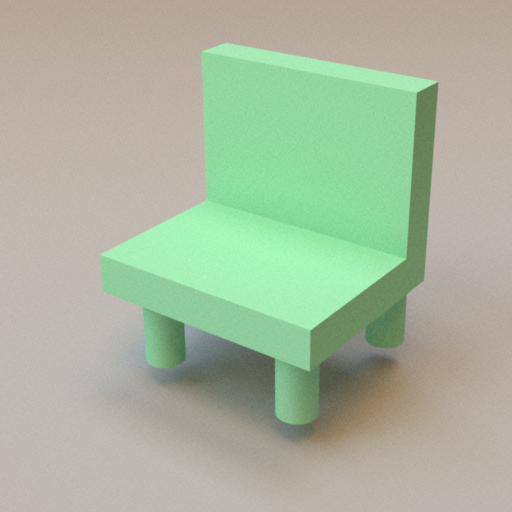}$\;$&
      \includegraphics[width = 2cm]{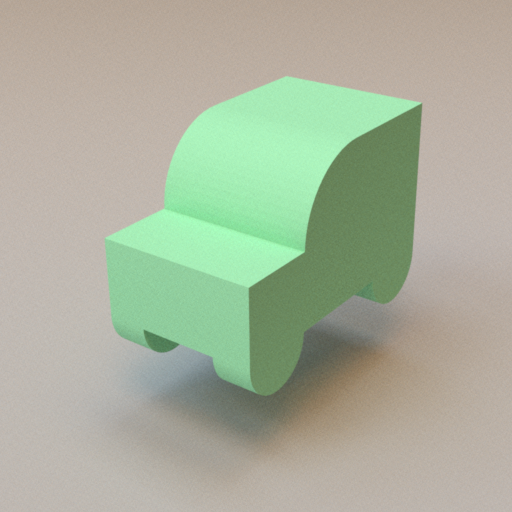}$\;$&
      \includegraphics[width = 2cm]{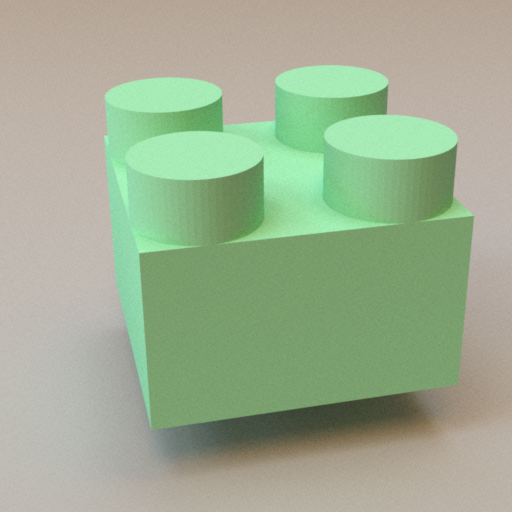}$\;$&
      \includegraphics[width = 2cm]{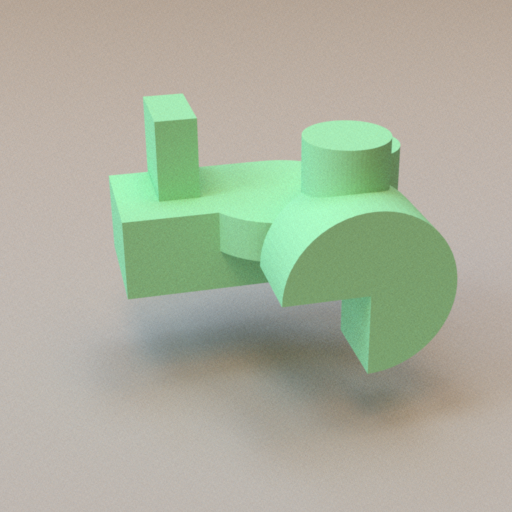}$\;$&
      \includegraphics[width = 2cm]{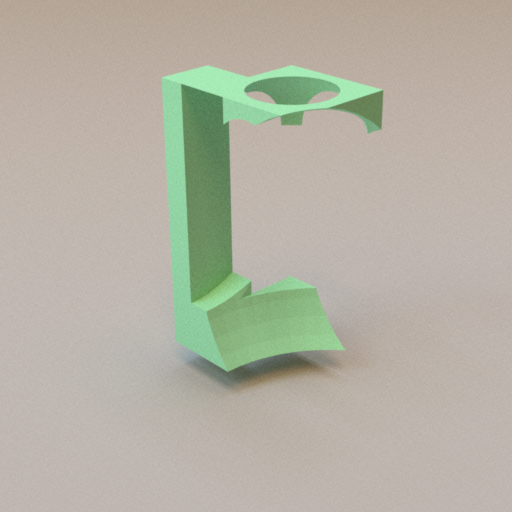}$\;$&
      \includegraphics[width = 2cm]{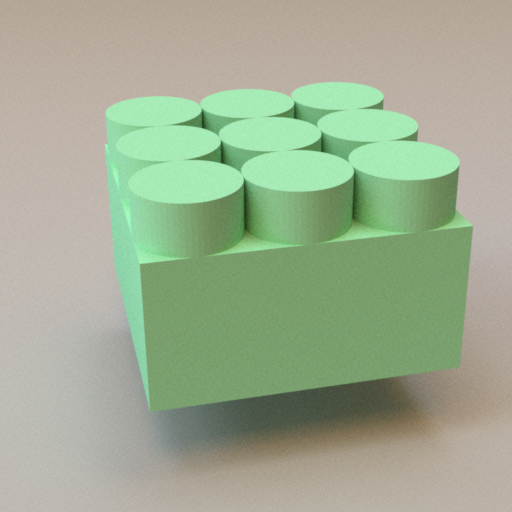}
    \end{tabular}
      \setlength{\tabcolsep}{6pt}
  \renewcommand{\arraystretch}{1}

  \end{tabular}
  \caption{Qualitative inverse CAD results. Top: Derendering random scenes vs. ablations and no-REPL baseline. Teal outlines show shape primitives in synthesized 2D programs. Bottom: Rerendering program inferred from voxels from novel viewpoints.}\label{graphics_qualitative}
\end{figure}

\textbf{Experimental evaluation}\quad 
We train our models on randomly generated scenes with up to 13 objects.
As a hard test of \emph{out-of-sample generalization} we test on randomly generated scenes with up to 30 or 20
objects for 2D and 3D, respectively.
Figure~\ref{graphics_quantitative} measures the quality of the best program found so far as a function of time,
where we measure the quality of a program by the intersection-over-union (IoU) with the spec.
Incorporating the value function proves important for both beam search and sampling methods such as SMC.
Given a large enough time budget the `no REPL' baseline is competitive with our ablated alternatives:
inference time is dominated by CNN evaluations, which occur at every step with a REPL,
but only once without it.
Qualitatively, an integrated policy, value network, and REPL yield programs closely matching the spec (Figure~\ref{graphics_qualitative}).
Together these components allow us to infer very long programs, despite a branching factor of $\approx$1.3 million per line of code: the largest programs we successfully infer\footnote{By ``successfully infer'' we mean IoU$\geq$0.9} go up to 19 lines of code/102 tokens for 3D
and 22 lines/107 tokens for 2D,
but the best-performing ablations fail to scale beyond 3 lines/19 tokens for 3D and 19 lines/87 tokens for 2D.



\subsection{String Editing Programs}
Learning programs that transform text is a classic program synthesis task~\cite{lau2001programming} made famous by the FlashFill system, which ships in Microsoft Excel~\cite{gulwani2011automating}.
We apply our framework to string editing programs using the RobustFill programming language~\cite{devlin2017robustfill},
which was designed for neural program synthesizers.

\textbf{REPL}\quad
Our formulation suggests modifications to the RobustFill language so that partial programs can be evaluated into a semantically coherent state (i.e. they execute and output something meaningful).
Along with edits to the original language, we designed and implemented a REPL for this domain, which, in addition to the original inputs and outputs, includes three additional features as part of the intermediate program state: The \textbf{committed} string maintains, for each example input, the output of the expressions synthesized so far. The \textbf{scratch} string maintains, for each example input, the partial results of the expression currently being synthesized until it is complete and ready to be added to the \emph{committed} string. Finally, the binary valued \textbf{mask} features indicate, for each character position, the possible locations on which transformations can occur. \todo{this might be a halfway explanation...} 

\begin{figure*}
  \centering
    \includegraphics[width=\textwidth]{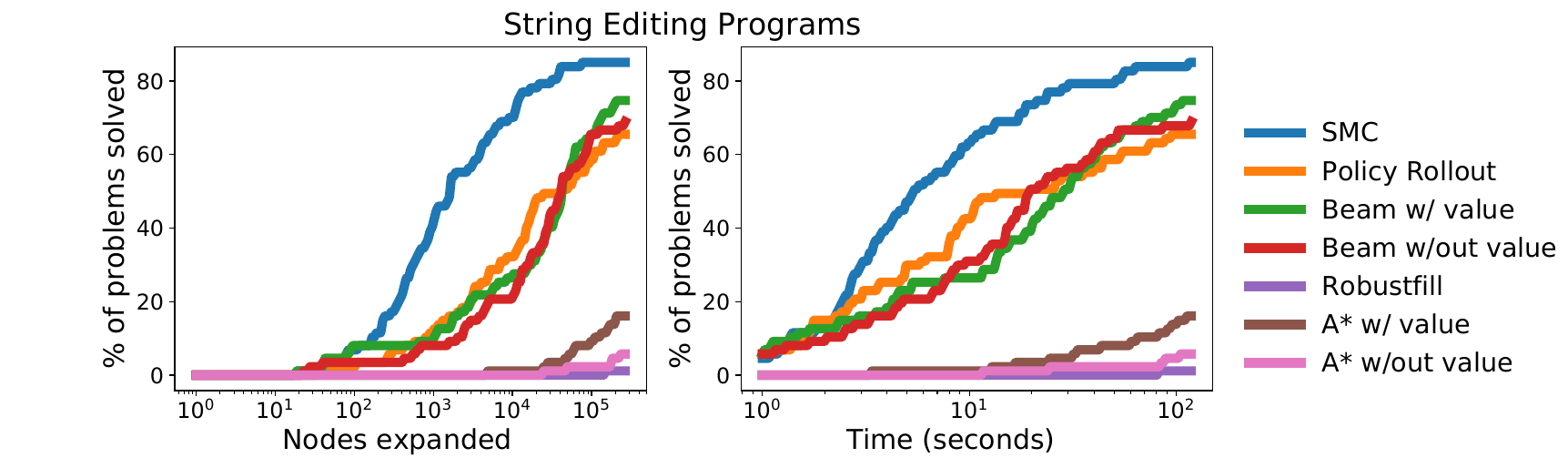}
  \caption{Results for String Editing tasks. Left: tasks solved vs number of nodes expanded. Right: tasks solved vs total time per task. Our SMC-based search algorithm solves more tasks using 10x fewer node expansions and less time than previous approaches. Note that x-axes are log scale.}
    \label{fig:robustfill}
\end{figure*}
\textbf{Experimental Evaluation}\quad
We trained our model and a reimplementation of RobustFill on string editing programs randomly sampled from the CFG. We originally tested on string editing programs from~\cite{nye2019learning} (comprising training tasks from~\cite{ellis2018learning} and the test corpus from~\cite{alur2016sygus}), but found performance was near ceiling for our model. We designed a more difficult dataset of 87 string editing problems from 34 templates comprising address, date/time, name, and movie review transformations. This dataset required synthesis of long and complex programs, making it harder for pure neural approaches such as RobustFill.

The performance of our model and baselines is plotted in Figure \ref{fig:robustfill}, and examples of best performing programs are shown in Figure \ref{tbl:rbtable}.
The value-guided SMC sampler leads to the highest overall number of correct programs, requiring less time and fewer nodes expanded compared to other inference techniques. We also observe that beam search attains higher overall performance with the value function than beam search without value. Our model demonstrates strong out-of sample generalization: Although it was trained on programs whose maximum length was 30 actions and average length approximately 8 actions, during test time we regularly achieved programs with 40 actions or more, representing a recovery of programs with description length greater than 350 bits.

\begin{figure}[]
\centering
  \setlength{\tabcolsep}{2pt}
  \renewcommand{\arraystretch}{1}
  \footnotesize
\begin{tabular}{|lll|l|lll|}
\cline{1-3} \cline{5-7}
\multicolumn{3}{|c|}{Spec:} &  & \multicolumn{3}{c|}{Spec:} \\
6/12/2003 & $\to$ & date: 12 mo: 6 year: 2003 &  & Dr Mary Jane  Lennon & $\to$ & Lennon, Mary Jane (Dr) \\
3/16/1997 & $\to$ & date: 16 mo: 3 year: 1997 &  & Mrs Isaac  McCormick & $\to$ & McCormick, Isaac (Mrs) \\ \cline{1-3} \cline{5-7} 
\multicolumn{3}{|c|}{Held out test instance:} &  & \multicolumn{3}{c|}{Held out test instance:} \\
12/8/2019 & $\to$ & date: 8 mo: 12 year: 2019 &  & Dr James Watson & $\to$ & Watson, James (Dr) \\ \cline{1-3} \cline{5-7} 
\multicolumn{3}{|c|}{Results:} &  & \multicolumn{3}{c|}{Results:} \\
\textbf{SMC (Ours)} & $\to$ & \textbf{date: 8 mo: 12 year: 2019} &  & \textbf{} & $\to$ & \textbf{Watson, James (Dr)} \\
Rollout & $\to$ & date: 8 mo: 1282019 &  &  & $\to$ & Watson, James \\
Beam w/value & $\to$ & date: 8 mo: 12 year:2019 &  &  & $\to$ & Watson, JamesDr \\
Beam & $\to$ & date: 8 mo: 12 year: &  &  & $\to$ & Watson, James ( \\
RobustFill & $\to$ & date:12/8/2019 &  &  & $\to$ & Dr James  Watson \\ \cline{1-3} \cline{5-7} 
\end{tabular}
  \setlength{\tabcolsep}{6pt}
  \renewcommand{\arraystretch}{1}
\caption{
Comparison of best programs on held-out inputs.
Best programs determined by Levenshtein distance of program outputs to spec outputs.
Leveraging the policy network, value network, and REPL-based execution guidance, SMC is able to consistently find programs with the desired behavior.}
\label{tbl:rbtable}
\end{figure}

\section{Discussion}
\paragraph{Related Work}
Within the program synthesis community, both text processing and graphics program synthesis have received considerable attention~\cite{gulwani2011automating}.
We are motivated by works such as InverseCSG~\cite{Du:2018:IAC:3272127.3275006}, CSGNet~\cite{sharma2018csgnet}, and RobustFill~\cite{devlin2017robustfill}, but our goal is not to solve a specific synthesis problem in isolation, but rather to push toward more general frameworks that demonstrate robustness across domains.

We draw inspiration from recent neural ``execution-guided synthesis'' approaches~\cite{zohar2018automatic,chen2018execution} which leverage partial evaluations of programs, similar to our use of a REPL. We build on this line of work by explicitly formalizing the task as an MDP,
which exposes a range of techniques from the RL and planning literatures. Our addition of a learned value network demonstrates marked improvements on methods that do not leverage such learned value networks.
Prior work~\cite{simmons2018program} combines tree search with $Q$-learning to synthesize small assembly programs,
but do not scale to large programs with extremely high branching factors, as we do (e.g., the $>40$ action-programs we synthesize for text editing or the $>$1.3 million branching factor per line of code in our 3D inverse CAD).





\paragraph{Outlook}

We have presented a framework for learning to write code combining two ideas: allowing the agent to explore a tree of possible solutions, and executing and assessing code as it gets written. This has largely been inspired by previous work on execution-guided program synthesis, value-guided tree search, and general behavior observed in how people write code.

An immediate future direction is to investigate programs with control flow like conditionals and loops. A 
Forth-style stack-based~\cite{rather1993evolution} language offer promising REPL-like representations of these control flow operators.
But more broadly, we are optimistic that many tools used by human programmers, like debuggers and profilers, can be reinterpreted and repurposed as modules of a program synthesis system. By integrating these tools into program synthesis systems, we believe we can design systems that write code more robustly and rapidly like people. 

\subsubsection*{Acknowledgments} We gratefully acknowledge many extended and productive conversations with Tao Du, Wojciech Matusik, and Siemens research. In addition to these conversations, Tao Du assisted by providing 3D ray tracing code, which we used when rerendering 3D programs. Work was supported by Siemens research, AFOSR award FA9550-16-1-0012 and the MIT-IBM Watson AI Lab. K. E. and M. N. are additionally supported by NSF graduate fellowships.

\bibliography{main}
\bibliographystyle{neurips_2019}
\newpage
\appendix
\section{Appendix}
\subsection{Graphics Programming Language}
\paragraph{CFG} The programs are generated by the CFG below:
\begin{lstlisting}[mathescape=true]
P -> P $+$ P | P $-$ P | S

# For 2D graphics  
S -> circle(radius=N, x=N, y=N)
   | quadrilateral(x0=N, y0=N,
                   x1=N, y1=N,
                   x2=N, y2=N,
                   x3=N, y3=N)

N -> [0 : 31 : 2]

# For 3D graphics
S -> sphere(radius=N, x=N, y=N, z=N)
   | cube(x0=N, y0=N, z0=N,
          x1=N, y1=N, z1=N)
   | cylinder(x0=N, y0=N, z0=N,
              x1=N, y1=N, z1=N, radius=N)
N -> [0 : 31 : 4]
\end{lstlisting}
In principle the 2D language admits arbitrary quadrilaterals. When generating synthetic training data we constrain the quadrilaterals to be take the form of rectangles rotated by 45 increments, although in principle one could permit arbitrary rotations by simply training a higher capacity network on more examples.

\subsection{String Editing Programming Language}

\paragraph{CFG} Our modified string editing language, based on~\cite{devlin2017robustfill} is defined as follows:
\begin{lstlisting}[mathescape=true]
Program P -> concat(E,.., E) $\text{(at most 6 E's)}$
Expression E -> F | N | N(N) 
            | N(F) | Const
Substring F -> Sub1(k) Sub2(k)  
             | Span1(r) Span2(i) Span3(y)
               Span4(r) Span5(i) Span6(y)
Nesting N -> GetToken1(t) GetToken2(i) 
            | ToCase(s) | GetUpTo(r)
            | GetFrom(r) | GetAll(t) 
            | GetFirst1(t) GetFirst2(i)
Regex r -> t | d
Type t -> Number | Word | AlphaNum
            | Digit | Char | AllCaps
            | Proper | Lower
Case s -> AllCaps | PropCase | Lower
Delimiter d ->  &,.?!@()[]%{}/#$\$$:;"'
Index i -> {-5 .. 6}
Boundary y -> Start | End
\end{lstlisting}

\subsection{Training details}

\paragraph{String Editing} We performed supervised pretraining for 24000 iterations with a batch size of 4000. We then performed REINFORCE for 12000 epochs with a batch size of 2000. Training took approximately two days with one p100 GPU.  We use the Adam optimizer wiht default settings.

Our Robustfill baseline was a re-implementation of the ``Attn-A" model from~\cite{devlin2017robustfill}. We implemented the ``DP-beam" feature, wherein during test-time beam search, partial programs which lead to an output string which is not a prefix of the desired output string are removed from the beam.
We trained for 50000 iterations with a batch size of 32. Training also took approximately two days with one p100 GPU.

\paragraph{2D/3D Graphics} We performed supervised pretraining with a batch size of 32, training on a random stream of CSG programs with up to 13 shapes, for approximately three days with one p100 GPU. We use the Adam optimizer wiht default settings. Over three days the 3D model saw approximately 1.7 million examples and the 2D model saw approximately 5 million examples.
We fine-tuned the policy using REINFORCE and trained the value network for approximately 5 days on one p100 GPU.
For each gradient step during this process we sampled $B_1 = 2$ random programs and performed $B_2 = 16$ rollouts for a total batch size of $B = B_1\times B_2 = 32$.
During reinforcement learning the 3D model saw approximately 0.25 million examples and the 2D model saw approximately 9 million examples.

For both domains, we performed no hyperparameter search. We expect that with some tuning, results could be marginally improved, but our goal is to design a general approach which is not sensitive to fine architectural details.

\subsection{Data and test-time details}

For both domains, we used a 2-minute timeout for testing for each problem, and repeatedly doubled the beam/number of particles until timeout is reached.

\paragraph{String Editing}
We originally tested on string editing programs from~\cite{nye2019learning} (comprising training tasks from~\cite{ellis2018learning} and the test corpus from~\cite{alur2016sygus}), but found our performance was near ceiling for our model (Figure \ref{fig:supportrb}). Thus, we designed our own dataset, as described in the main text. Generation code for this dataset can be found in our supplement, in the file \texttt{generate\_test\_robust.py}.

\begin{figure}
\includegraphics[width = \textwidth]{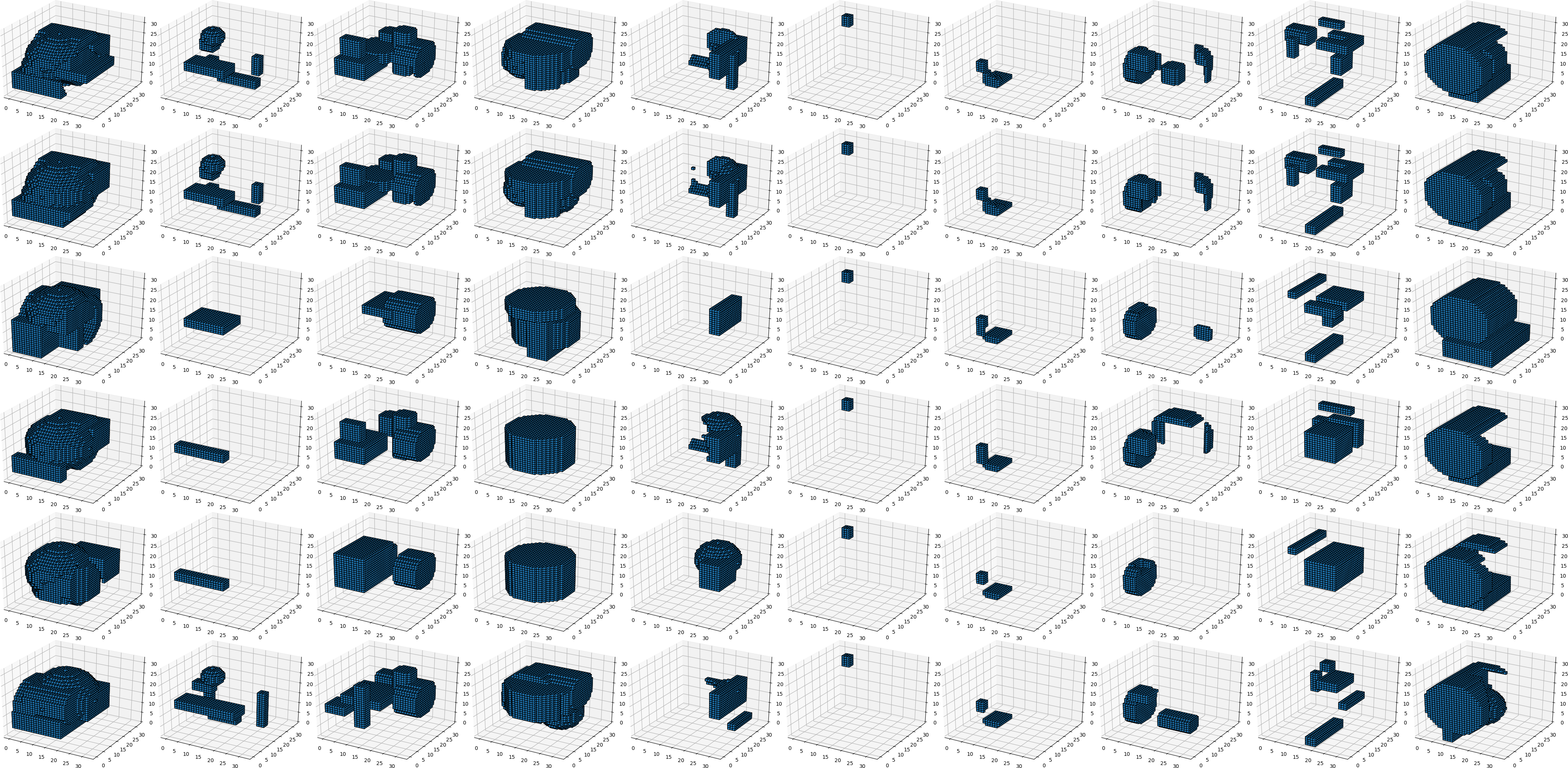}
\includegraphics[width = \textwidth]{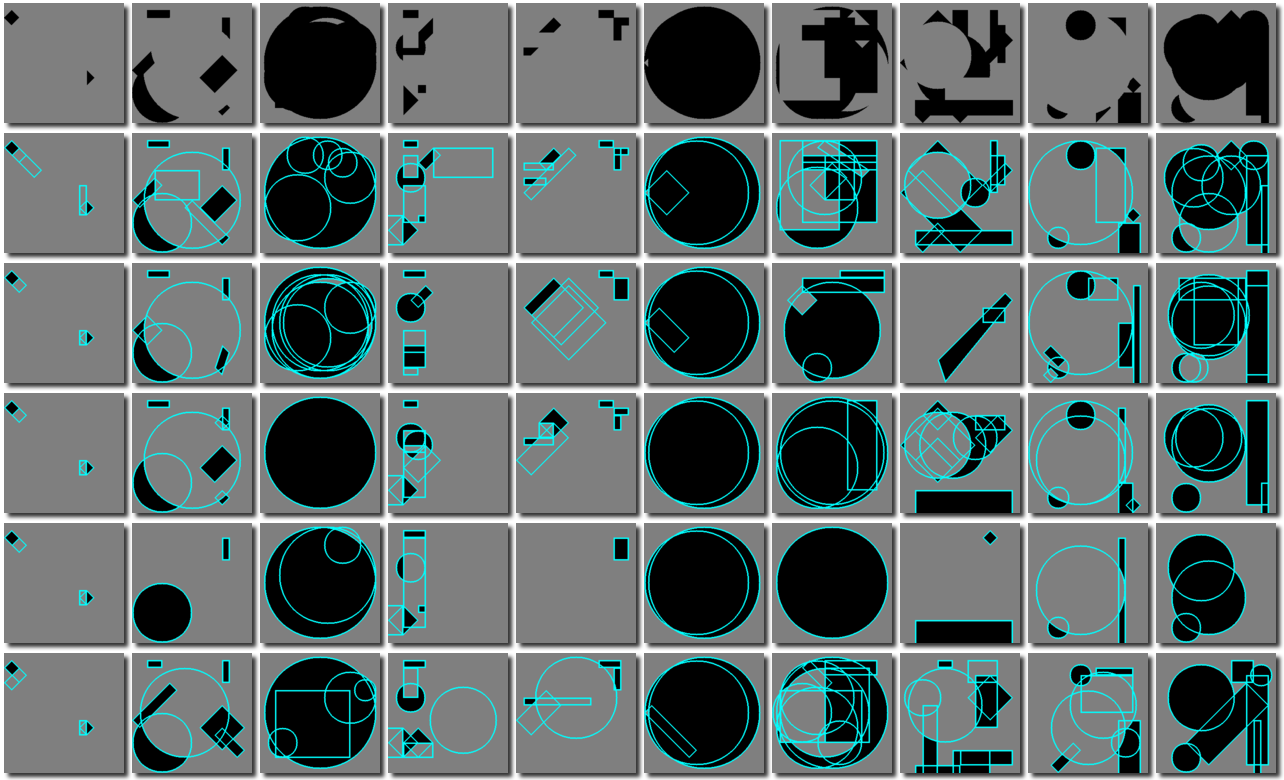}
                       \caption{Top: Spec. Bottom rows (in order): SMC, policy rollouts, beam search with value, beam search without value, no REPL}\label{3dpixel_montage}
\end{figure}

\paragraph{2D/3D Graphics} We generate a scene with up to $k$ objects by sampling a number between 1 to $k$,
and then sampling a random CSG tree with that many objects.
We then remove any subtrees that do not affect the final render (e.g., subtracting pixels from empty space).
Our held-out test set is created by sampling 30 random scenes with up to $k=20$ objects for 3D and $k=30$ objects for 2D.
Running \texttt{python driver.py demo --maxShapes 30} using the attached supplemental source code will generate example random scenes.
Figure~\ref{3dpixel_montage} illustrates ten random 3-D/2-D scenes and contrasts different model outputs.

\subsection{Architecture details}


\subsubsection{String Editing}
For this domain, our neural architecture involves encoding each example state separately and pooling into a hidden state, which is used to decode the next action.
To encode each example, we learn an embedding vector of size 20 for each character and apply it to each position in the input string, output string, committed string, and scratch string. For each character position, we concatenate these embedding vectors, additionally concatenating the values of the masks for that spatial position. We then perform a 1-d convolution with kernel size 5 across the character positions. Following \cite{zohar2018automatic}, we concatenate the vectors for all the character positions, and pass this through a dense block with 10 layers and a growth rate of 128 to produce a hidden vector for a single example. We perform an average pooling on the  hidden vector for each example. We then concatenate the resulting vector with a 32-dim embedding of the previous action and apply a linear layer, which results in the final state embedding, from which we decode the next action.
Our value network is identical, except the final layer instead decodes a value.

\subsubsection{Inverse CAD}
The policy is a CNN followed by a pointer network which decodes into the next line of code.
A pointer network~\cite{vinyals2015pointer} is an RNN that uses a differentiable attention mechanism to emit pointers, or indices, into a set of objects.
Here the pointers index into the set of partial programs $pp$ in the current state, which is necessary for the union and difference operators.
Because the CNN takes as input the current REPL state -- which may have a variable number of objects in scope -- we encode each object with a separate CNN and sum their activations, i.e. a `Deep Set' encoding~\cite{zaheer2017deep}. The value function is an additional `head' to the pooled CNN activations.

Concretely the neural architecture has a \emph{spec encoder}, which is a CNN inputting a single image,
as well as a \emph{canvas encoder}, which is a CNN inputting a single canvas in the REPL state,
alongside the spec, as a two-channel image.
The canvas encoder output activations are summed and concatenated with the spec encoder output activations
to give an embedding of the state:
\begin{align}
\text{stateEncoding}(spec,pp) = \text{specEncoder}(spec)\otimes\sum_{p\in pp}\text{canvasEncoder(spec,\eval{p})}
\end{align}
for $W_1$, $W_2$ weight matrices.

For the policy we pass the state encoding to a pointer network,
implemented using a GRU with 512 hidden units and one layer,
which predicts the next line of code.
To attend to canvases $p\in pp$, we use the output of the canvas encoder as the `key' for the attention mechanism.

For the value function we passed the state in coding to a MLP with 512 hidden units w/ a hidden ReLU activation,
and finally apply a negated `soft plus' activation to the output to ensure that the logits output by the value network is nonpositive:
\begin{align}
v(spec,pp) &= -\text{SoftPlus}(W_2\text{ReLU}(W_1\text{stateEncoding}(spec,pp)))\\
\text{SoftPlus}(x)& = \log (1 + e^x)
\end{align}
\noindent\textbf{2-D CNN architecture:} The 2D spec encoder and canvas encoder both take as input $64\times 64$ images,
passed through 4 layers of 3x3 convolution,
with ReLU activations after each layer and 2x2 max pooling  after the first two layers.
The first 3 layers have 32 hidden channels and the output layer has 16 output channels.

\noindent\textbf{3-D CNN architecture:} The 3D spec encoder and canvas encoder both take as input $32\times 32\times 32$ voxel arrays,
passed through 3 layers of 3x3 convolution,
with ReLU activations after each layer and 4x4 max pooling  after the first layer.
The first 2 layers have 32 hidden channels and the output layer has 16 output channels.

\noindent\textbf{No REPL baseline:} Our ``No REPL''  baselines using the same CNN architecture to take as input the \emph{spec}, and then use the same pointer network architecture to decode into the program,
with the sole difference that, rather than attend over the CNN encodings of the objects in scope (which are hidden from this baseline),
the pointer network attends over the hidden states produced at the time when each previously constructed object was brought into scope.

\subsection{String editing additional results}

\begin{figure*}
  \centering
    \includegraphics[width=\textwidth]{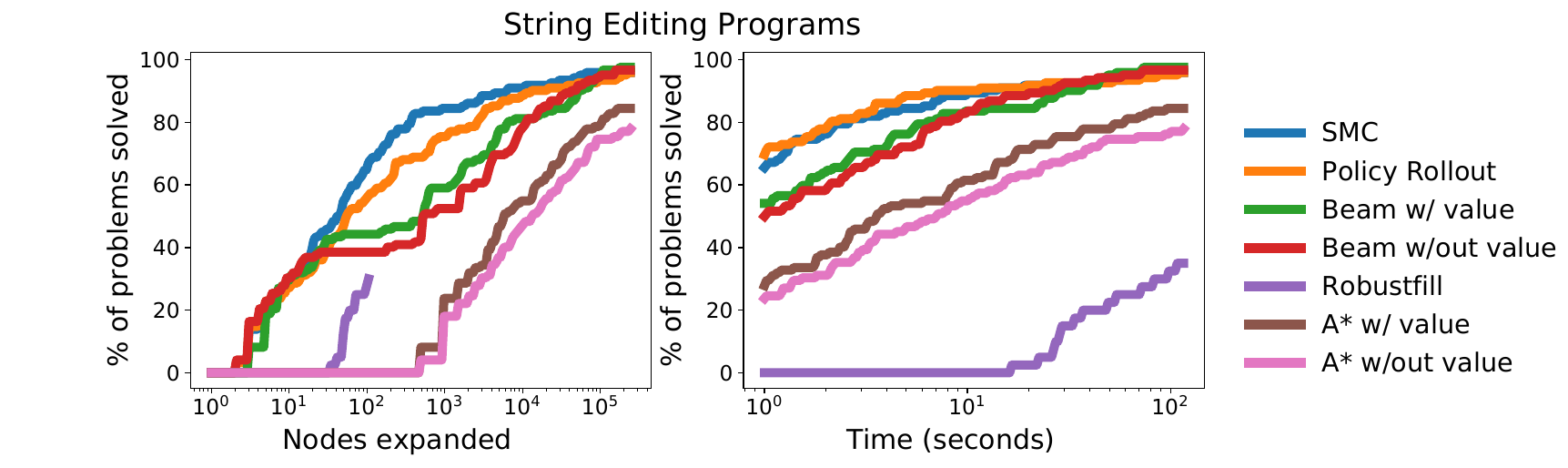}
  \caption{Results for String Editing tasks on dataset from \cite{nye2019learning}. Left: tasks solved vs number of nodes expanded. Right: tasks solved vs total time per task. Note that x-axes are log scale.}
    \label{fig:supportrb}
\end{figure*}

Figure \ref{fig:supportrb} shows results on the string editing dataset from \cite{nye2019learning}.

\begin{figure}
\fbox{
\parbox{\textwidth}
{
inputs:\\
('3/16/1997', '4/17/1986', '6/12/2003', '4/23/1997')\\
outputs:\\
('date: 16 mo: 3 year: 1997', 'date: 17 mo: 4 year: 1986', 'date: 12 mo: 6 year: 2003', 'date: 23 mo: 4 year: 1997')\\
Const(d), Commit, Const(a), Commit, Const(t), Commit, Const(e), Commit, Const(:), Commit, Const( ), Commit, Replace1(/), Replace2( ), GetToken1(Number), GetToken2(1),\\
Commit, Const( ), Commit, Const(m), Commit, Const(o), Commit, Const(:), Commit, Const( ), Commit, GetUpTo(Number), Commit, Const( ), Commit, Const(y), Commit,\\ Const(e), Commit, Const(a), Commit, Const(r), Commit, Const(:), Commit, Const( ), Commit, GetFrom(/), Commit
}
}

\fbox{
\parbox{\textwidth}
{
inputs:\\
('April 19, 2:45 PM', 'July 5, 8:42 PM', 'July 13, 3:35 PM', 'May 24, 10:22 PM')\\
outputs:\\
('April 19, approx. 2 PM', 'July 5, approx. 8 PM', 'July 13, approx. 3 PM', 'May 24, approx. 10 PM')\\
GetUpTo( ), Commit, GetFirst1(Number), GetFirst2(-3), Commit, Const(,), Commit, Const( ), Commit, Const(a), Commit, Const(p), Commit, Const(p), Commit, Const(r),\\
Commit, Const(o), Commit, Const(x), Commit, Const(.), Commit, Const( ), Commit, GetFrom(,), GetFirst1(Digit), GetFirst2(3), GetFirst1(Digit), GetFirst2(-3), Commit,\\
Const( ), Commit, Const(P), Commit, Const(M), Commit
}
}

\fbox{
\parbox{\textwidth}
{
inputs:\\
('cell: 322-594-9310', 'home: 190-776-2770', 'home: 224-078-7398', 'cell: 125-961-0607')\\
outputs:\\
('(322) 5949310 (cell)', '(190) 7762770 (home)', '(224) 0787398 (home)', '(125) 9610607 (cell)')\\
Const((), Commit, ToCase(Proper), GetFirst1(Number), GetFirst2(1), GetFirst1(Char), GetFirst2(2), Commit, Const()), Commit, Const( ), Commit, GetFirst1(Number),\\
GetFirst2(5), GetFirst1(Char), GetFirst2(-2), GetToken1(Char), GetToken2(3), Commit, SubStr1(-16), SubStr2(17), GetFirst1(Number), GetFirst2(4), GetToken1(Char),\\
GetToken2(-5), Commit, GetFirst1(Number), GetFirst2(5), GetToken1(Char), GetToken2(-5), Commit, GetToken1(Number), GetToken2(2), Commit, Const( ), Commit,\\
Const((), Commit, GetUpTo(-), GetUpTo(Word), Commit, Const()), Commit
}}

\fbox{
\parbox{\textwidth}
{
inputs:\\
('(137) 544 1718', '(582) 431 0370', '(010) 738 6792', '(389) 820 9649')\\
outputs\\
('area code: 137, num: 5441718', 'area code: 582, num: 4310370', 'area code: 010, num: 7386792', 'area code: 389, num: 8209649')\\
Const(a), Commit, Const(r), Commit, Const(e), Commit, Const(a), Commit, Const( ), Commit, Const(c), Commit, Const(o), Commit, Const(d), Commit, Const(e), Commit,\\
Const(:), Commit, Const( ), Commit, GetFirst1(Number), GetFirst2(0), Commit, Const(,), Commit, Const( ), Commit, Const(n), Commit, Const(u), Commit, Const(m),\\
Commit, Const(:), Commit, Const( ), Commit, GetFrom()), GetFirst1(Number), GetFirst2(2), Commit
}}

\label{fig: rbsamples}
\caption{Examples of long programs inferred by our system in the string editing domain.}
\end{figure}

\end{document}